# Increasing Flexibility of Combined Heat and Power Systems through Optimal Dispatch with Variable Mass Flow

Xin Qin, *Student Member*, *IEEE*, Ye Guo, Senior *Member*, *IEEE*, Xinwei Shen, *Member*, *IEEE*, Hongbin Sun, *Fellow IEEE*

*Abstract*—In combined heat and power systems, varying mass flow can better make use of the heating system's inertia to increase the flexibility of electric power systems. This is challenging, however, due to integer variables and bilinear constraints in existing optimal dispatch models. In this paper, we incorporate an improved heat pipeline model to eliminate complexity from integer variables without compromise on accuracy. Subsequently, the resulting optimal dispatch model with bilinear constraints is solved by the proposed modified Generalized Benders Decomposition method, which decomposes the optimal dispatch model into a convex sub-problem with the fixed mass flow and a simple upper-level problem searching for the optimal mass flow. Comparisons with existing benchmarks show that the proposed method can achieve lower operation costs with outstanding computational efficiency.

## I. INTRODUCTION

### A. Motivation

Combined heat and power systems are widely deployed all over the world. For example, 30%-50% of total electric power is generated by combined heat and power units in northern Europe [1] and northern China [2]. Such a large-scale application is not only driven by the high fuel efficiency but also the complementary properties of the two energy sectors: The electric power system requires real-time power balance whereas the temperature in a heating system has much higher inertia. Thus, if operated properly, the heating system can serve as a storage for the electric power system, which may in turn provide alternative heat sources to heating systems. Therefore, it is of crucial importance to exploit such a complementary property with all possible adjustment means.

In the optimal dispatch of combined heat and power systems, temperature and mass flow are the two most important types of control variables in heating systems [4][5]. There are two widely-used adjustment means in heating systems: varying temperature adjustment (with fixed mass flow) and varying both temperature and mass flow adjustment. Compared with the former, the latter can increase the adjustable range of heat power and improve system flexibility because temperature is varying in only a small range, which is much smaller than the range of varying mass flow [20]. In practice, the supervisory control and data acquisition (SCADA) system can automatically control different devices to adjust the supply temperature and mass flow of heat sources to the setting values [6]. With these advantages, it is crucial to adjust both mass flow and temperature in heating systems.

### B. Related Works

Adjusting both mass flow and temperature is challenging because it brings integer variables and bilinear constraints to the optimization model, making the optimization model difficult to solve.

Integer variables are used to describe the time delay of the pipeline heat transmission when mass flow is variable. For example, paper [16] models the optimal dispatch with variable mass flow as an Mixed-Integer Nonlinear Program (MINLP). Based on the MINLP model in [16], Vesterlund [17] and Wang [18], et al. consider multiple heat sources in the heating system. However, the MINLP models in [16]-[18] are extremely hard to solve because of incorporating both integer variables and bilinear constraints. To reduce complexity from integers, different simplification manners have been proposed. For instance, paper [21] uses binary variables to substitute the integer variables in the MINLP models of [16]-[18]. Papers [19] and [20] eliminate integer variables in optimal dispatch models by ignoring the heat dynamic process. However, the manners in [19]-[21] over-simplify heating system models, which can lead to economic inefficiency and make the optimization problems infeasible to solve.

In order to deal with bilinear constraints, two types of approaches have been developed to make the optimal dispatch with variable mass flow tractable. One is linearizing heat flow models based on Energy Hub [7] and Ohm's law [8]. However, the linearized models cannot fully consider the transmission limits and the flexibility from heat inertia. The other approach is to apply convexification techniques to this problem. Li [10] and Lin [11] et al. formulate convex optimal dispatch programs by fixing mass flow to use pipeline inertia to accommodate renewables. Moreover, papers [12] and [13] consider the integration of multiple distributed electric and heat sources for more generation flexibility. However, the methods in [10]-[13] cannot fully make use of heat inertia because the heat mass flow is fixed, which restricts the further improvement of system flexibility.

The appeal of better flexibility and accuracy motivates other researchers to propose the third approach: Developing advanced solution methods to deal with the bilienarity caused by varying mass flow. Two kinds of solution methods, data-driven methods and model-based methods, have been developed. The former use historical data to produce dispatch strategies [14][15]. Although the online optimization of these data-driven methods may be efficient, their results may suffer from problems of interpretability and reliability, especially under different conditions. Heuristic algorithms can be embedded in data-driven methods to improve reliability and accuracy. For example, papers [16]-[18] propose heuristic iterative methods to solve optimal dispatch models with bilinear constraints. Unfortunately, due to the bilinearity, heuristic methods still have problems in convergence and interpretability. Recently, model-based methods are developed to address the interpretability and convergence problems of data-driven methods. For instance, papers [7] and [20] adopt convex relaxation for bilinear constraints, but the heating network is over-simplified. Authors of [21] have proposed a modified Generalized Benders Decomposition (GBD) method to deal with the bilinear constraints in the optimal dispatch model. However, the GBD method in [21] has simplifications and approximations in the solution procedure, which impedes the security and optimality.

In summary, there is still no perfect solutions to deal with complexities brought by varying mass flow, which hinders further developments of combined heat and power systems.

### C. Summary of Contributions

In this paper, we address the challenges of integer variables and bilinear constraints in the optimal dispatch for combined heat and power systems with variable mass flow. We improve the heat pipeline model which describes heat inertia of pipelines under variable mass flow: The integer variables for reflecting time delays in existing pipeline models are eliminated by a series of time-correlated bilinear pipeline heat transmission equations without compromise on accuracy. After that, the optimal dispatch model with variable mass flow only has bilinear constraints without integers.

Furthermore, to deal with bilinearity, we propose a modified GBD method that decomposes the optimal dispatch model with bilinear constraints into a convex lower-level sub-problem and a simple upper level master problem. The sub-problem is formulated by fixing mass flow to obtain a convex problem, which is efficient to solve. The convex sub-problem provides the master problem with the gradient direction to update mass flow or the cutting planes to remove infeasible mass flow regions. The master problem uses the gradient descent method to search for better mass flow with lower overall costs based on the information from the sub-problems. By the iteration between the sub-problem and the master problem, the proposed method can effectively reduce the total operation costs with satisfactory computational efficiency. Compared with data-driven methods and heuristic methods, the proposed method is more reliable and interpretable.

The remainder of this paper is organized as follows. In Section II, the heat power flow model is formulated without integer variables. In Section III, we present the optimal dispatch model for combined heat and power systems and decompose it into a convex sub-problem with fixed mass flow and an upper-level problem updating mass flow. The solution method for the lower and upper level problems is proposed in Section V. In section V, numerical tests have been done to compare the proposed method with existing benchmarks.

## II. POWER FLOW MODEL

To give a clear picture of the combined heat and power system, we introduce its physical model and mathematical power flow model in this section. In this paper, we discuss the regional combined heat and power system in the size of a town, so the electric power system and the heating system can be jointly dispatched.

### A. Physical Model

As presented in Fig. 1, the electric power system and the heating system are coupled through energy sources including thermal generator, combined heat and power (CHP) unit, electric boiler, tie-line connected to the main grid, etc.

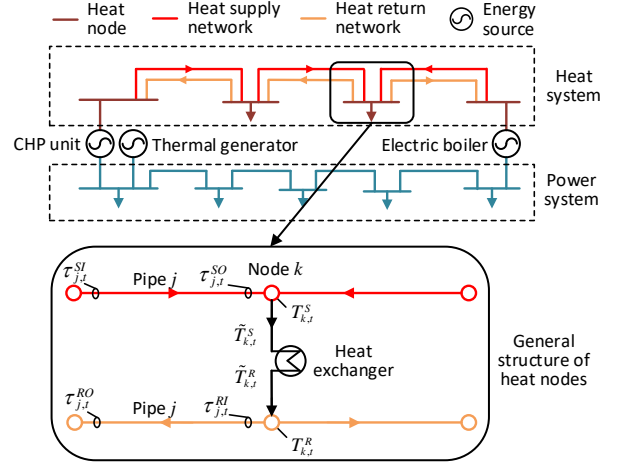

Fig. 1. General structure of combined heat and power system.

As shown in Fig.1, the heating system consists of heat nodes, supply network (in red line), and return network (in yellow line). In the heating system, heat is produced at source nodes and carried by the high-temperature water in the supply network. Load nodes use heat exchangers to obtain heat from high-temperature water in the supply network and release the low-temperature water to the return network. The low-temperature water flows from load nodes to the source nodes in the return network and is reheated by the heat sources. For details of mass flow circulation in the heating system, see [33].

### B. Heat Power Flow Model

In the heating system, it is assumed that 1) the heat supply network and the heat return network are radial, respectively, and 2) the mass flow is nonnegative. These assumptions hold in most practical heating systems and are widely adopted in the literature [16][20][21].

For clarity, in the following parts, we use $\tau$ and $T$ to denote the pipe temperature and the node temperature, respectively. The superscripts $S$ and $R$ indicate the variables in supply and return networks, respectively. Superscript $t$ indicates the variables at period $t$.

### 1) Heat Node Model

As shown in Fig. 1, the node obtains heat power using heat exchangers between the supply network and the return network. Therefore, the heat node $k$ has 4 related variables: $T_{k,t}^S$ and $T_{k,t}^R$ indicate the node supply and return temperatures of node $k$ at period $t$, respectively. Similarly, $\tilde{T}_{i,t}^S$ and $\tilde{T}_{i,t}^R$ indicate exchanger supply and return temperatures, respectively. Node temperature is the mixing of exchanger temperature and pipe temperature:

$$\left(m_{k,t}^n + \sum_{j \in P^S \cap I(k)} m_{j,t}\right) T_{k,t}^S = \left(m_{k,t}^n \tilde{T}_{k,t}^S\right) + \left(\sum_{j \in P^S \cap I(k)} m_{j,t} \tau_{j,t}^{SO}\right), \quad (1)$$
$$\forall k \in H, \ t = 1, 2, \ldots, N$$

$$\left(m_{k,t}^n + \sum_{j \in P^R \cap I(k)} m_{j,t}\right) T_{k,t}^R = \left(m_{k,t}^n \tilde{T}_{k,t}^R\right) + \left(\sum_{j \in P^R \cap I(k)} m_{j,t} \tau_{j,t}^{RO}\right), \quad (2)$$
$$\forall k \in H, \ t = 1, 2, \ldots, N$$

where $m_{k,t}^n$ is the node mass flow of node $k$, and $m_{j,t}$ is the pipe mass flow of pipeline $j$. Scalars $\tau_{j,t}^{SO}$ and $\tau_{j,t}^{RO}$ are pipe outlet temperatures in heat supply and return networks, respectively. $N$ is the number of total time periods. Sets $P^S$ and $P^R$ indicate the set of pipelines in supply and return networks, respectively. Set $H = H_L \cup H_G$ is the set of heat nodes, where $H_G$ and $H_L$ are the sets of heat source nodes and load nodes, respectively. Set $I(k)$ is the set of pipelines injecting into node $k$. It is noticed, for load nodes, $m_{k,t}^n$ in (1) equals to 0; For source nodes, $m_{k,t}^n$ in (2) equals to 0.

The node heat power $h_{k,t}$ is calculated by:

$$h_{k,t} = c_p m_{k,t}^n (\tilde{T}_{k,t}^S - \tilde{T}_{k,t}^R) \quad \forall k \in H, \ t = 1, 2, \ldots, N, \quad (3)$$

where $c_p$ is the heat capacity of water. In the supply network, for load nodes $T_{k,t}^S = \tilde{T}_{k,t}^S$, and for source nodes $T_{k,t}^S$ is calculated by (1). Similarly, in the return network, for load nodes $T_{k,t}^R$ is calculated by (2), and for source nodes $T_{k,t}^R = \tilde{T}_{k,t}^R$.

The node mass flow satisfies the hydraulic Kirchhoff's law: The difference of the pipe mass flow injecting into and leaving from a node equals to the node mass flow.

$$\boldsymbol{A} \boldsymbol{m}_t = \boldsymbol{m}_t^n \quad t = 1, 2, \ldots, N, \quad (4)$$

where $\boldsymbol{m}_t$ and $\boldsymbol{m}_t^n$ are pipe and node mass flow vectors, respectively. $\boldsymbol{A}$ is the node-branch incidence matrix defined as [19], in which

$$A_{k,j} = \begin{cases} +1, & \text{the mass flow of pipe } j \text{ comes into node } k \\ -1, & \text{the mass flow of pipe } j \text{ leaves from node } k \\ 0, & \text{no connection from pipe } j \text{ to node } k \end{cases}$$

Since the heating network is assumed to be radial, for clarity, in the following parts, we use (4) to replace the node mass flow with the pipe mass flow.

*2) Heat Pipeline Model*

In this paper, we innovate the method of describing heat pipeline dynamics: Pipeline model with integer variables in [16]-[18] are substituted by the time-correlated and space-correlated bilinear pipeline heat transmission equations (5)-(6), where (5) is for supply network and (6) is for return network. For the derivation process, see Appendix. As verified in [22]-[24], (5)-(6) can accurately describe the pipeline heat dynamic process. It is noticed, the equation in [22]-[24] are used for the heating system simulation, where pipe temperatures $\tau_j^S$ and $\tau_j^R$ as well as mass flow $m_{j,t}$ of pipe $j$ are given values. But in (5)-(6) $\tau_j^S$, $\tau_j^R$ and $m_{j,t}$ are decision variables.

$$(a_j m_{j,t} + b_j) \tau_j^S(i,t) = c_j \tau_j^S(i,t-1) + b_j m_{j,t} \tau_j^S(i-1,t) + d_j, \quad \forall j \in P^S, \ i = 1, 2, \cdots S_j, \ t = 1, 2, \ldots, N \quad (5)$$

$$(a_j m_{j,t} + b_j) \tau_j^R(i,t) = c_j \tau_j^R(i,t-1) + b_j m_{j,t} \tau_j^R(i-1,t) + d_j, \quad \forall j \in P^R, \ i = 1, 2, \cdots S_j, \ t = 1, 2, \ldots, N \quad (6)$$

where $\tau_j^S(i,t)$ and $\tau_j^R(i,t)$ are pipe temperatures at segment $i$ from the pipe inlet point at period $t$. $S_j = \lceil x_j / \Delta x \rceil$ indicates the segment section number of pipe $j$, where $x_j$ is the length of pipe $j$, and $\Delta x$ is the given segment length. Scalars $a_j$-$d_j$ are coefficients related to the parameters of pipe $j$:

$$a_j = \frac{1}{\Delta t} + \frac{1}{\rho c_p A_j R_j}, \ b_j = \frac{1}{\Delta x \rho A_j}, \ c_j = \frac{1}{\Delta t}, \ d_j = \frac{T_{j,t}^a}{\rho c_p A_j R_j},$$

where $\rho$ is the density of water. Scalar $\Delta t$ is the time interval between two time periods. Scalars $A_j$ and $R_j$ are the cross-sectional area and the thermal conductive coefficient of pipe $j$, respectively. Scalar $T_{j,t}^a$ is the ambient temperature of pipe $j$ at period $t$.

Moreover, there are boundary limits for (5) and (6) including $\tau_j^S(0,t) = \tau_{j,t}^{SI}$ and $\tau_j^S(S_j,t) = \tau_{j,t}^{SO}$ in the supply network and $\tau_j^R(0,t) = \tau_{j,t}^{RI}$ and $\tau_j^R(S_j,t) = \tau_{j,t}^{RO}$ in the return network, where $\tau_{j,t}^{SI}$ and $\tau_{j,t}^{RI}$ denote pipe inlet temperatures in heat supply and return networks, respectively.

The pipe inlet temperature is equal to the temperature of its connecting node:

$$\tau_{j,t}^{SI} = T_{k,t}^S \quad j \in P^S \cap L(k), \ k \in H, \ t = 1, 2, \ldots, N, \quad (7)$$
$$\tau_{j,t}^{RI} = T_{k,t}^R \quad j \in P^R \cap L(k), \ k \in H, \ t = 1, 2, \ldots, N, \quad (8)$$

where $L(k)$ is the set of pipelines leaving from node $k$.

*C. Electric Power Flow Model*

In the electric power system, the DC power flow model is adopted[1]. The real-time electric power balance is required between the generation side and the load side:

$$\sum_{i \in E} p_{i,t} = \sum_{i \in E} d_{i,t} \quad t = 1, 2, \ldots, N, \quad (9)$$

where $p_{i,t}$ denotes the electric power generation of bus $i$ at period $t$. If a bus does not have energy sources, then its $p_{i,t} = 0$. Scalar $d_{i,t}$ is the electric load demand. Set $E$ denote the set of electric buses.

The transmission line power flow $l_{i,t}$ of line $i$ is calculated by:

$$l_{i,t} = \sum_{j \in E} SF_{i,j} \cdot (p_{j,t} - d_{j,t}) \quad \forall i \in \varGamma, \ t = 1, 2, \ldots, N, \quad (10)$$

where $SF_{i,j}$ indicates the shift factor of bus $j$ to line $i$. Set $\varGamma$ is the set of electric power lines.

## III. OPTIMIZATION MODEL

Now we have the electric and heat power flow models. Next, in this section, we formulate the optimization model for the combined heat and power dispatch and propose the decomposition strategy to transform the optimization model with bilinear constraints to a solvable form.

*A. Optimization Model*

*1) Electric Power System Constraints*

The electric line power must satisfy the thermal limitation:

$$-\overline{l}_{i,t} \leq \sum_{j \in E} SF_{i,j} \cdot (p_{j,t} - d_{j,t}) \leq \overline{l}_{i,t} \quad \forall i \in L, \ t = 1, 2, \ldots, N, \quad (11)$$

where $\overline{l}_{i,t}$ is the line power limit of line $i$ at period $t$.

Moreover, the electric power flow equations (9)-(10) are included in the optimization model as equality constraints.

---
[1] Here we use the most well-known DC power flow model. Other linear power flow models like distflow model [32] can be adopted without changing the solution method in Section V.

*2) Heating System Constraints*

The pipe mass flow $m_{j,t}$ should satisfy:

$$\underline{m}_{j,t} \le m_{j,t} \le \overline{m}_{j,t} \quad \forall j \in P^S \cup P^R,\ t=1,2,...,N,\quad (12)$$

$$m_{j,t} \ge 0 \quad \forall j \in P^S \cup P^R,\ t=1,2,...,N,\quad (13)$$

where $\underline{m}_{j,t}$ and $\overline{m}_{j,t}$ are the lower and upper limits of mass flow of pipe $j$ at period $t$, which not only consider the pipe mass flow limits but also the node mass flow limits. The (13) can guarantee the pipe mass flow is nonnegative when multiple heat sources are integrated.

To prevent the heat pipeline inertia from being exhausted, the generated heat energy is required to be no less than the load heat energy within scheduling periods:

$$\sum_{t=1}^{N}\left(\sum_{k \in H_G} h_{k,t} - \sum_{k \in H_L} h_{k,t}\right) \ge 0.\quad (14)$$

Moreover, the heat power flow equations (1)-(8) are included in the optimization model as equality constraints.

*3) Energy Source Constraints*

The feasible regions of different kinds of electric and heat sources are described by polytopes [10][16][29]:

$$B_{k,i} p_{i,t} + K_{k,i} h_{i,t} \le v_{k,i} \quad \forall i \in G,\ t=1,2,...,N,\quad (15)$$

where $B_{k,i}$, $K_{k,i}$ and $v_{k,i}$ are coefficients of the $k$ th boundary of the feasible operating region of energy source $i$. Set $G$ is the set of energy sources. For example,

1) As shown in Fig. 2 (a), if a source generates electricity and heat simultaneously such as a CHP unit, its polytope is in the first quadrant, where $p_{i,t} \ge 0$ and $h_{i,t} \ge 0$. Similarly, the polytope of an electric boiler resembles that of the CHP unit, but is in the fourth quadrant.
2) As shown in Fig. 2 (b), if a source only generates electricity such as a thermal generator, the coefficient $K_{k,i}$ related to heat power output is zero, where $p_{i,t} \ge 0$ and $h_{i,t} = 0$.
3) As shown in Fig. 2 (c), if a source only generates heat such as a natural gas boiler, the coefficient $B_{k,i}$ related to electric power output is zero, where $p_{i,t} = 0$ and $h_{i,t} \ge 0$.

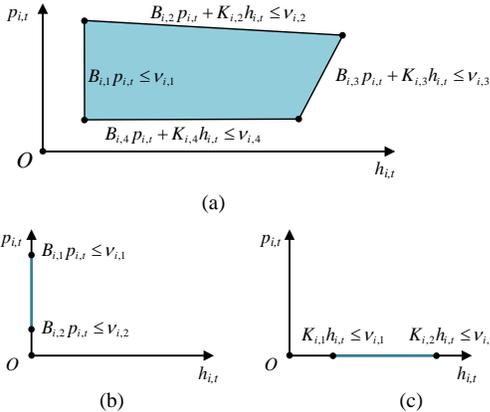

Fig. 2. The feasible regions of (a) CHP units, (b) thermal generators, and (c) natural gas boilers.

The energy source must also satisfy ramping constraints: The increment or decrement of the source power outputs within a single period should not exceed the ramping capacities:

$$D_{e,i} \cdot \Delta t \le p_{i,t} - p_{i,t-1} \le U_{e,i} \cdot \Delta t \quad \forall i \in G,\ t=2,3,...,N,\quad (16)$$

$$D_{h,i} \cdot \Delta t \le h_{i,t} - h_{i,t-1} \le U_{h,i} \cdot \Delta t \quad \forall i \in G,\ t=2,3,...,N,\quad (17)$$

where $D_{e,i}$ and $U_{e,i}$ are the downward and upward electric ramping capacities of source $i$, respectively. $D_{h,i}$ and $U_{h,i}$ are the downward and upward heat ramping capacities of source $i$.

*4) Objective Function*

The objective function of the optimal dispatch is to minimize the operation costs of all energy sources in all time periods:

$$\min f = \sum_{i \in G}\sum_{t=1}^{T} C_{i,t}(p_{i,t}, h_{i,t}),\quad (18)$$

where $C_{i,t}$ is the cost function of energy source $i$ at period $t$, which is expressed using a quadratic function of electricity and heat productions and is assumed to be convex in this paper [26]:

$$C_{i,t} = \eta_{i,t,0} + \eta_{i,t,1} p_{i,t} + \eta_{i,t,2} p_{i,t}^2 + \eta_{i,t,3} h_{i,t} + \eta_{i,t,4} h_{i,t}^2 + \eta_{i,t,5} p_{i,t} h_{i,t},\quad (19)$$

where $\eta_{i,t,0}$-$\eta_{i,t,5}$ are the cost coefficients of energy source $i$ at period $t$, which are given by generation costs and electricity price. For example, for thermal generators which only generate electricity, coefficients of heat-related terms are zero. For electric boilers, electricity-related coefficients are negative and heat-related coefficients are positive.

### B. Model Analysis and Decomposition

*1) Model Generalization*

For clarity, here we summarize the optimization model in Section III-A in a succinct way. Since the node mass flow is substituted by the pipe mass flow using (4), we let $\boldsymbol{m} = [\boldsymbol{m}_t]$ be the vector of pipe mass flow, where the dimension of $\boldsymbol{m}$ is $(n_P \cdot N) \times 1$, and $n_P$ is the number of heat pipelines. Then we let $\boldsymbol{x} = [\boldsymbol{p}, \boldsymbol{h}, \boldsymbol{l}, \tilde{\boldsymbol{T}}^S, \tilde{\boldsymbol{T}}^R, \boldsymbol{T}^S, \boldsymbol{T}^R, \boldsymbol{\tau}^{SI}, \boldsymbol{\tau}^{SO}, \boldsymbol{\tau}^{RI}, \boldsymbol{\tau}^{RO}, \boldsymbol{\tau}^S, \boldsymbol{\tau}^R]$ denote the matrix of other decision variables. After the two steps, the optimization model of combined heat and power dispatch with variable mass flow can be written as:

$$\min_{\boldsymbol{x},\boldsymbol{m}} f(\boldsymbol{x}),$$
$$s.t.\ h_1(\boldsymbol{x},\boldsymbol{m}) = 0,\ h_2(\boldsymbol{x}) = \boldsymbol{\alpha}_0^T \boldsymbol{x} + \boldsymbol{\beta}_0 = 0,\quad (20)$$
$$g_1(\boldsymbol{x}) = \boldsymbol{\alpha}_1^T \boldsymbol{x} + \boldsymbol{\beta}_1 \le 0,\ g_2(\boldsymbol{m}) = \boldsymbol{\alpha}_2^T \boldsymbol{m} + \boldsymbol{\beta}_2 \le 0,$$

where the objective function $f(\boldsymbol{x})$ denotes the $f$ in (18). $\boldsymbol{\alpha}_0$-$\boldsymbol{\alpha}_2$ and $\boldsymbol{\beta}_0$-$\boldsymbol{\beta}_2$ are coefficient matrices. The meanings of constraints are shown in Table I:

TABLE I
The Meaning of Constraints in (20)

| Constraint | Meaning | Equation number |
|---|---|---|
| $h_1$ | Nonlinear coupling equality constraints between $\boldsymbol{x}$ and $\boldsymbol{m}$ | (1)-(3), (5)-(6) |
| $h_2$ | Linear equality constraints on $\boldsymbol{x}$ only | (7)-(10), boundary limits of (5)-(6) |
| $g_1$ | Linear inequality constraints on $\boldsymbol{x}$ only | (11), (14)-(17) |
| $g_2$ | Linear constraints on $\boldsymbol{m}$ only | (12)-(13) |

Since equation (9) is applied to eliminate the node mass flow, we do not have equality constraints on $\boldsymbol{m}$ only.

*2) Model Decomposition*

The challenge of solving the optimization model (20) is that it is a nonconvex program with bilinear constraints $h_1(\boldsymbol{x},\boldsymbol{m}) = 0$. Although the problem (20) is nonconvex, if $\boldsymbol{m}$ is fixed, it will become a standard convex programming, which is convenient to solve.

Based on the idea of GBD, we treat mass flow $\boldsymbol{m}$ as the coupling variable. Thus, as shown in Fig.3, the problem (20) can be decomposed into a convex lower-level sub-problem with fixed $m$ and an upper-level master problem which optimizes $m$:

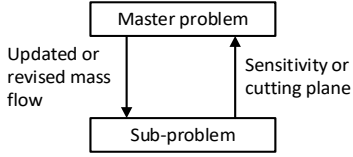

Fig. 3. The architecture of model decomposition.

In what follows, we formulate the sub-problem and the master problem respectively [21][25].

*a) Convex Sub-Problem*

The sub-problem (21) is constructed by fixing mass flow:

$$\begin{aligned}&\min_{x} f(x),\\ &s.t.\ h_1(x,m^k)=0,\ h_2(x)=\alpha_0^T x+\beta_0=0,\\ &\qquad g_1(x)=\alpha_1^T x+\beta_1 \le 0,\end{aligned} \quad (21)$$

where $m^k$ indicates the mass flow $m$ at $k$ th iteration. It is noticed the problem (21) is a standard convex problem because 1) the objective function is convex, and 2) all constraints are linear.

Formulated at $m^k$, the sub-problem provides the local information around $m^k$ for the master problem to search for a better $m$.

*b) Master Problem*

The master problem is a mapping of the optimal cost function with respect to mass flow $m$ which uses the information from sub-problems to search for better $m$ in its parameter space:

$$\begin{aligned}&\min_{m} J^*(m),\\ &m \in M \cap FC,\end{aligned} \quad (22)$$

where $J^*(m)$ is the optimal cost function of $m$. $M$ indicates the original parameter space of $m$ constructed by $g_2(m)$, and $FC$ indicates cutting planes.

## IV. SOLUTION METHOD

We propose the modified GBD method to solve the optimization model (21) by iterations between the sub-problem and the master problem. The lower-level sub-problem is convex, which can be efficiently solved by existing solvers like CPLEX. Solving the upper-level master problem (22) is challenging because we do not have any closed-form expression of $J^*(m)$. In this section, the gradient method is proposed to solve the master problem at the neighborhood of $m^k$.

Solving the lower-level sub-problem results in two cases: Feasible sub-problem and infeasible sub-problem.

### A. Feasible Sub-Problem

If the sub-problem is feasible, the master problem updates $m$ based on the sensitivity calculated by the sub-problem. First, the sensitivity of the optimal cost function with respect to the mass flow $m$ is calculated using the envelope theorem:

$$\frac{\partial J^*(m^k)}{\partial m_{i,t}} = \frac{\partial f^*(x)}{\partial m_{i,t}}\bigg|_{x=x^k,m=m^k} = \frac{\partial L(x,m)}{\partial m_{i,t}}\bigg|_{x=x^k,m=m^k} \quad (23)$$

where $f^*(x)$ is the optimal cost function, and $L(x,m)$ is the Lagrangian function of the sub-problem. $x^k$ is the variable $x$ at $k$ th iteration.

Second, the master problem updates the mass flow $m$ by moving along the anti-gradient direction using (24) [29]. Here the objective function $J^*(m)$ is approximated by the first-order gradient (23), and the constraints are considered by the projection matrix $P^k$.

$$m^{k+1} = m^k - (\alpha^k P^k)\frac{\partial J^*(m^k)}{\partial m}, \quad (24)$$

where $\alpha^k$ is the step size at $k$ th iteration:

$$\alpha^k = \frac{\gamma J^*(x^k)}{\left(-P^k \frac{\partial J^*(m^k)}{\partial m}\right)^T \left(\frac{\partial J^*(m^k)}{\partial m}\right)}, \quad (25)$$

in which $\gamma$ is the desired reduction rate set from 10% to 50%. The gradient term in (25) is provided by (23). Matrix $P^k$ is the projection matrix at $k$ th iteration which incorporates possible active boundary constraints of $g_2(m)$ and cutting planes:

$$P^k = I - H_A^k \left((H_A^k)^T H_A^k\right)^{-1} (H_A^k)^T, \quad (26)$$

where $H_A^k$ indicates the matrix of active constraints, for more details, see [30].

### B. Infeasible Sub-Problem

If the sub-problem is infeasible, we revise the mass flow to its feasible region by removing the infeasible region of $m$ from the original parameter space $M$. To this end, first we construct the relaxed sub-problem and solve it:

$$\begin{aligned}&\min_{x,s}\sum_{i} s_i,\\ &s.t.\ \bar{\lambda}:\ h_1(x,m^k)=0,\\ &\qquad \bar{\mu}:\ g_1(x)=\alpha_1^T x+\beta_1 \le s,\\ &\qquad h_2(x)=\alpha_0^T x+\beta_0=0,\end{aligned} \quad (27)$$

where $s_i$ is the slack variable for the $i$ th inequality constraint, and $s$ is the vector of slack variables. $\bar{\lambda}$ and $\bar{\mu}$ are dual variable matrices for $h_1$ and $g_1$, respectively.

Second, the cutting plane is generated based on Outer Approximation [25][27] to cut infeasible regions of $m$ because the problem (28) is a convex program with all constraints linear:

$$(\bar{\lambda}^k)^T [\nabla_m h_1(x^k,m^k)^T (m-m^k)] + (\bar{\mu}^k)^T g_1(x^k) \le 0, \quad (28)$$

where $\bar{\lambda}^k$ and $\bar{\mu}^k$ are the values of $\bar{\lambda}$ and $\bar{\mu}$ at $k$ th iteration, respectively.

When solving the relaxed sub-problem (27), we can obtain the values of $\bar{\lambda}^k$, $\bar{\mu}^k$, $\nabla_m h_1(x^k,m^k)^T$, and $g_1(x^k)$. Thus, inequality equation (28) is linear with $m$ as variables only, which defines a cutting plane removing the infeasible region of $m$ from the original parameter space $M$. Given by Outer Approximation, the cutting plane (27) can accelerate the calculation by reducing iteration times for infeasible sub-problems [28].

Third, the master problem revises $m$ according to cutting planes, where the revised mass flow denotes the intersection of the gradient direction and the cutting plane:

$$m^{k+1} = m^r - \beta^k \frac{\partial J^*(m^r)}{\partial m}, \quad (29)$$

where $r$ indicates the last successful iteration, and $\beta^k$ indicates the step size for revision:

$$\beta^k = \frac{\left(\bar{\lambda}^k\right)^T \left[\nabla_m h_1(x^k, m^k)^T (m^r - m^k)\right] + \left(\bar{\mu}^k\right)^T g_1(x^k)}{\left(\bar{\lambda}^k\right)^T \left[\nabla_m h_1(x^k, m^k)^T\right] \left(\frac{\partial f(x^r, m^r)}{\partial m}\right)}. \quad (30)$$

### C. Example of Solution Process

For clarity, an example of the solution process is illustrated in Fig. 4. First, from $m^i$ to $m^{i+1}$, the master problem updates $m$ based on the gradient with projection according to (24), because if the projection is not considered, $m$ will break the constraints of $g_2(m)$. Second, in the process from $m^{i+1}$ to $m^{i+2}$, the master problem updates $m$ according to (24) in the gradient direction without projection, where $P^{i+1} = I$. Third, after the process from $m^{i+2}$ to $m^{i+3}$, the sub-problem is infeasible with $m^{i+3}$. Thus, a cutting plane is generated according to (28) for the master problem. Forth, the process from $m^{i+3}$ to $m^{i+4}$ indicates the process (29) of revising $m$ based on the cutting plane. The $m^{i+4}$ is the intersection point of the gradient direction and the cutting plane. Last, the process from $m^{i+4}$ to $m^{i+5}$ finds the local optimum $m^{i+5}$; Therefore, the iteration stops at $m^{i+5}$.

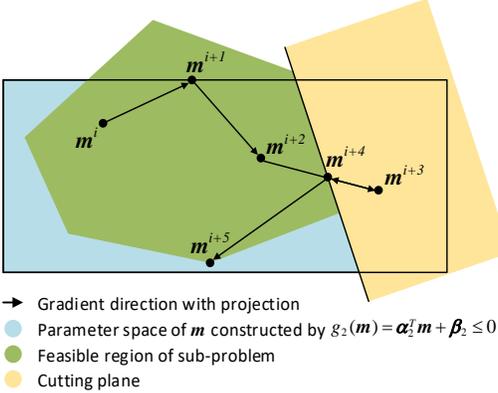

- → Gradient direction with projection
- ● $m^i$
- ▨ Parameter space of $m$ constructed by $g_2(m) = \alpha_2^T m + \beta_2 \leq 0$
- ▨ Feasible region of sub-problem
- ▨ Cutting plane

Fig. 4. The example of the solution process.

The convergence criterion $\sigma^k$ is defined as:

$$\sigma^k = \left|\frac{(J^*)^k - (J^*)^{k-1}}{(J^*)^1}\right|. \quad (31)$$

Given $\delta$ as the maximum convergence tolerance, if $\sigma^k \leq \delta$, the iteration will stop at $k$ th iteration.

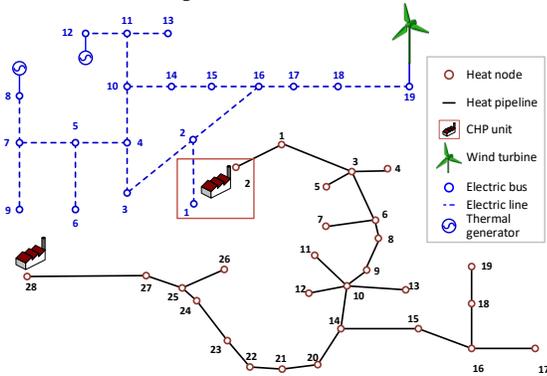

Fig. 5. The topology of the combined heat and power system in a city in Northeast China.

## V. CASE STUDIES

In case studies, the first case compares the proposed method with the separate dispatch and the fixed mass flow dispatch to illustrate why varying mass flow helps improve the system's flexibility. The second case compares different optimal dispatch methods with variable mass flow to demonstrate the proposed method's advantages in optimality and convergence.

The case simulations are performed on a laptop with a 2.80 GHz CPU and the 16 GB memory. Programs are coded using Matlab, and YALMIP is used as a socket between Matlab and solvers CPLEX for the convex sub-problem (22) and (25). The calculation convergences if the $\sigma^k \leq 1 \times 10^{-4}$ or the sub-problem is infeasible after 3 consecutive iterations.

### A. Flexibility from Varying Mass Flow

This case is carried out to demonstrate the benefits of varying mass flow in the combined heat and power dispatch. The proposed method is compared with two widely-used methods: the separate dispatch of electricity and heat (separate method) and the combined heat and power dispatch with fixed mass flow [13] (fixed flow method).

As shown in Fig. 5, the simulation is based on a real system in Northeast China with a 19-bus transmission-level electric power system and a 28-node heating system. The heating system has two sources, and the total pipeline length is about 16 km. The detailed topology and data are uploaded on [31].

TABLE II
Comparison of Overall Costs and Renewable Curtailment

| Methods | Overall costs ($) | Renewable curtailment rate | Renewable curtailment penalty ($) |
|---|---|---|---|
| Separate dispatch | $3.547 \times 10^6$ | 13.98% | $7.611 \times 10^4$ |
| Fixed flow method | $2.979 \times 10^6$ | 1.27% | $6.927 \times 10^3$ |
| Proposed method | $2.666 \times 10^6$ | 0 | 0 |

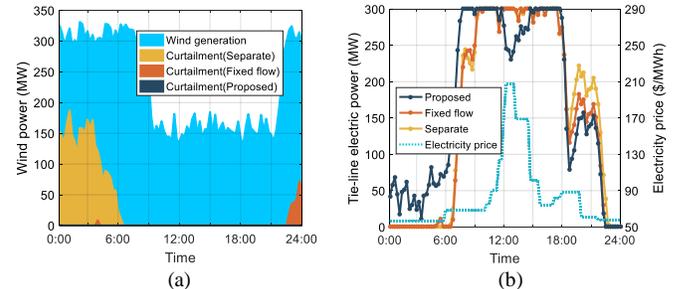

(a)     (b)

Fig. 6. (a) The renewable curtailment and (b) the electricity purchase from the main grid.

As shown in Table II, the two combined dispatch methods, i.e., the fixed flow method and the proposed method, have reduced 16.01% and 24.84% of the overall costs as well as 90.92% and 100% renewable curtailments compared with the separate dispatch, respectively. This significant improvement comes from the flexibility provided by the heating system: In the separate dispatch, the electric power system dispatch only considers the value of heat load power rather than the model of the heating system. As a result, heat generation should strictly follow the heat load curve as shown in Fig. 7 (c) with no flexibility from the heating system. However, as shown in Fig. 7 (a) and (b), when jointly dispatching the two energy systems, the heat pipeline network can serve as a storage for the electric power system to reduce costs. For example, the pipeline network discharges to satisfy the heat load during 0:00-6:00 as shown in Fig. 7 (a) and (b), which can enlarge the adjustment ranges of CHP electric and heat power outputs and yields to less wind power curtailment as shown in Fig. 6 (a).

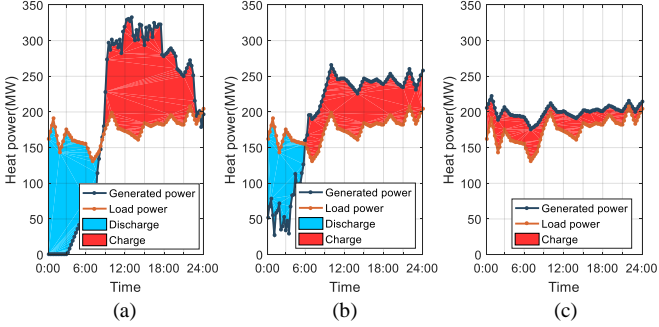

Fig. 7. The heat power generation and heat load of (a) the proposed method, (b) the fixed flow method, and (c) separate dispatch.

On the other hand, varying mass flow can further increase system flexibility: The proposed method has reduced 10.51% overall operation costs and eliminated all renewable curtailment compared with the fixed flow method. As shown in Fig. 7, compared with the fixed flow method, the proposed method allows CHP units to generate more electricity and heat within 9:00-20:00 because varying mass flow makes better use of heat inertia in the pipeline network. Therefore, as presented in Fig. 6 (b), the system operator can purchase less high-price electricity from the main grid within 10:30-14:00. Moreover, by varying mass flow, the proposed method eliminates all the wind curtailment compared with the fixed flow method and realizes 100% renewable accommodation.

In terms of calculation efficiency, the solver time of the proposed method is only 15.45s, which can be used in day-ahead dispatch or intra-day dispatch. Briefly, varying mass flow in the optimal dispatch of combined heat and power systems can better make use of heat inertia to increase system flexibility, which contributes to lower renewable curtailment and operation costs.

### B. Comparison with Existing Methods

Based on the test system in Fig. 8, we compare different variable mass flow dispatch methods in terms of convergence and optimality.

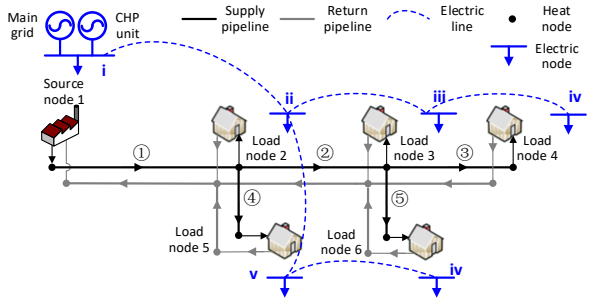

Fig. 8. The topology of the 6-node test system.

As shown in Table III, both the method in [16] and the direct method fail to converge under the given condition. The two methods' divergence has a similar reason: Both methods directly use heuristic methods and commercial solvers to solve the nonconvex optimization models with bilinear constraints, which cannot guarantee convergence for this kind of optimization programs. Meanwhile, each method has its own drawbacks which may lead to the divergence: 1) The MINLP model in [16] has integer variables as well as bilinear constraints, which is extremely difficult to solve; 2) The interior point method adopted by IPOPT cannot guarantee the convergence when dealing with large-scale bilinear constraints.

TABLE III
Comparison of Different Dispatch Methods with Variable Mass Flow

| Methods | Overall costs ($) | Solver CPU time (s) | YALMIP time (s) |
|---|---|---|---|
| The method in [16] | Fail to converge | 49.49 | 6.81 |
| The proposed model solved by the method in [21] | 4.929×10⁵ | 0.87 | 90.7 |
| The proposed model directly solved by IPOPT (direct method) | Fail to converge | 194.48 | 63.25 |
| The proposed method | 4.798×10⁵ | 5.48 | 609.28 |

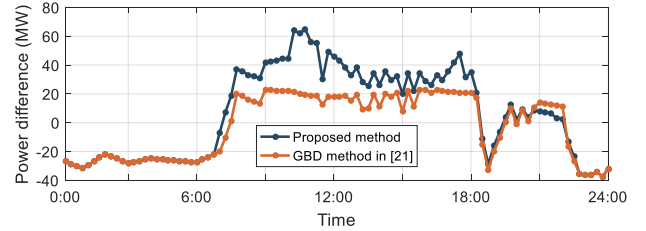

Fig. 9. The heat power difference between the generation side and the load side.

Moreover, as shown in the grey blocks in Table III, both the proposed method and the method in [21] successfully overcome the divergence problem of the other two methods, while the overall operation costs of the proposed method are 2.66% lower than the method in [21]. The reason is that by better adjustment of mass flow, the proposed method enlarges the pipeline storage capacity compared with the method in [21] as shown in Fig. 9, which further improves the adjustable range of CHP electric and heat power outputs; Therefore, as shown in Fig. 10 (a) CHP units can generate more electricity to reduce purchasing high-price electricity from the grid during 10:00-18:00; As shown in the light blue area in Fig. 10 (b), the proposed method can also improve the renewable accommodation from 6:00 to10:00 by better using heat inertia.

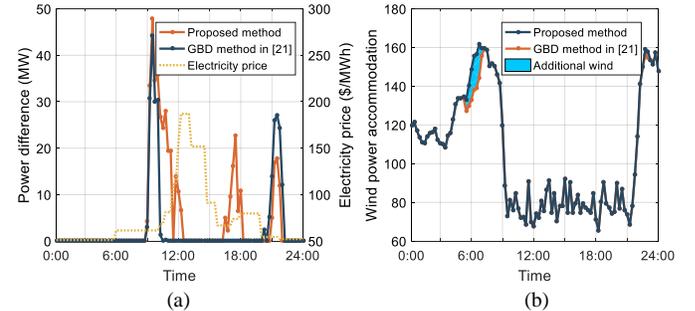

Fig. 10. The (a) electricity purchase from the main grid and (b) wind power accommodation.

Now we discuss why the proposed method can better guarantee optimality than the method in [21]: Since the optimization model (20) is nonconvex, when the sub-problem is feasible the optimal cut given by the method in [21] is an over-conservative approximation of the accurate cutting plane. As a result, the master problem may remove $m$ with better optimality based on the approximated optimal cuts. By contrast, the proposed method does not approximate the cutting planes but uses the step-by-step update to search for $m$ with lower overall costs, which avoids the over-conservative approximation in [21] and can better ensure optimality.

As shown in Table III, the proposed method has high computational efficiency, which spends 5.48 s solving the 96-period day-ahead optimal dispatch model. Compared with the method in [21], the proposed method reduces 1.31×10⁴ $ costs at the expense of increasing 4.61s CPU time, which is acceptable for the day-ahead economic dispatch. It is noticed for research convenience and program generality, the YALMIP is used as a socket between Matlab and solvers, which consumes additional time to load and transform constraints. But if we directly use solvers, this time consumption will not appear.

In summary, the proposed method can better ensure optimality by relieving the over-conservative approximation in existing research and overcome the divergence problem by eliminating integer variables and designing the proper decomposition mechanism.

## VI. CONCLUSION

In this paper, the flexibility of combined heat and power systems is increased through the optimal dispatch with variable mass flow, which makes better use of heat inertia. To eliminate complexity from integer variables in optimal dispatch models, we innovate to use bilinear pipeline heat transmission equations which accurately describe the heat pipeline dynamic process. To deal with bilinear constraints, we propose a modified GBD method that decomposes the nonconvex optimal dispatch model with bilinear constraints into a convex sub-problem with the fixed mass flow and a simple master problem using the gradient descent method to search for better mass flow. Compared with existing combined heat and power dispatch methods with variable mass flow, the proposed method can address the divergence problem of heuristic methods and better guarantee optimality by overcoming over-conservative approximation. Compared to optimal dispatch methods with fixed mass flow, the proposed method enlarges the heat pipeline storage capacity by varying mass flow, resulting in lower overall costs and renewable curtailment.